\documentclass[twocolumn,secnumroman,amssymb, nobibnotes, aps, prd]{revtex4}

\usepackage{graphicx}
\usepackage{dcolumn}
\usepackage{bm}
\usepackage{color}
\begin{document}

%\preprint{}

\title{Microwave-induced Hall resistance in bilayer electron systems}

\author{S. Wiedmann,$^{1,2,3}$ G. M. Gusev,$^4$ O. E. Raichev,$^5$ S. Kr\"{a}mer,$^2$ A. K. Bakarov,$^6$ and J. C. Portal$^{2,3}$}
\affiliation{$^1$Radboud University Nijmegen, Institute for Molecules and Materials, High Field Magnet Laboratory, Toernooiveld 7, 6525 ED Nijmegen, the Netherlands} 
\affiliation{$^2$Laboratoire National des Champs Magn\'{e}tiques Intenses, CNRS-UJF-UPS-INSA, 38042 Grenoble, France} 
\affiliation{$^3$INSA Toulouse, 31077 Toulouse Cedex 4, France} 
\affiliation{$^4$Instituto de F\'{\i}sica da Universidade de S\~ao Paulo, CP 66318, S\~ao Paulo, SP,Brazil}
\affiliation{$^5$Institute of Semiconductor Physics, NAS of Ukraine, Prospekt Nauki 41, 03028, Kiev, Ukraine} 
\affiliation{$^6$Institute of Semiconductor Physics, Novosibirsk 630090, Russia}

\date{\today}

\begin{abstract}
The influence of microwave irradiation on dissipative and Hall resistance in high-quality bilayer 
electron systems is investigated experimentally. We observe a deviation from odd symmetry under 
magnetic field reversal in the microwave-induced Hall resistance $\Delta R_{xy}$ whereas the 
dissipative resistance $\Delta R_{xx}$ obeys even symmetry. Studies of $\Delta R_{xy}$ as 
a function of the microwave electric field and polarization exhibit a strong and non-trivial 
power and polarization dependence. The obtained results are discussed in connection to 
existing theoretical models of microwave-induced photoconductivity. 
\end{abstract}

\pacs{73.40.-c, 73.43.-f, 73.21.-b}

\maketitle

\section{Introduction}

In the last decade, it has been found that an external ac field (microwaves - MW's) 
causes the appearance of microwave-induced resistance oscillations (MIRO's) \cite{1}
which evolve into zero-resistance states (ZRS) for high-quality two-dimensional electron 
systems (2DES) in the presence of a perpendicular magnetic field \cite{2}. MIRO's and ZRS 
occur in dissipative resistance but are not accompanied by plateaus in Hall resistance
as for the integer quantum Hall effect \cite{3}. MIRO periodicity is governed by the ratio of 
radiation frequency $\omega$ to cyclotron frequency $\omega_{c}=eB/m$, where $m$ is 
the effective mass of the electrons. In theory, it is currently assumed that these
oscillating phenomena can be explained by mechanisms originating from the scattering-assisted 
electron transitions between different Landau levels (LLs) in the presence of microwave excitation. 
The two main competing microscopic mechanisms for oscillating photoresistance are the ``displacement" 
mechanism which accounts for spatial displacement of electrons along the applied dc field under 
scattering-assisted microwave absorption \cite{4,5}, and the ``inelastic" mechanism, owing to an 
oscillatory contribution to the isotropic part of the electron distribution function \cite{6,7}. 
Such a consideration describes the periodicity and phase of MIRO's observed in experiments.

Recently, it has been demonstrated that MW-induced phenomena in 2DES are not restricted to 
single-layer 2DES. Microwave-induced resistance oscillations have been found in bilayer and 
trilayer systems \cite{8,9}, and high-mobility bilayers with two occupied 2D subbands 
exhibit ZRS \cite{10}. The specific features in magnetoresistance in bilayers and multilayers 
are caused by an interference of magneto-intersubband (MIS) oscillations \cite{11} with MIRO's, 
when MW irradiation enhances, suppresses, or inverses the MIS oscillations. 

Apart from dissipative resistance, one can ask if and how MW irradiation
does affect Hall resistance since it was first a surprise, see Mani \textit{et al.} in 
Ref. \cite{2}, that Hall effect seemed to be unaffected by microwaves. Subsequent experiments 
on high-quality single-layer 2DES have shown weak MW-induced oscillations in Hall resistance 
\cite{12,13}. The MW-induced Hall resistance $\Delta R_{xy}=R_{xy}-R_{xy}^{(0)}$, where $R_{xy}^{(0)}$ 
is the dark Hall resistance, depends on MW power and follows $1/B$-periodicity of photoresponse 
in dissipative resistance. The observed {\em odd symmetry} under field reversal, 
$\Delta R_{xy}(B) = - \Delta R_{xy}(-B)$, is preserved under increasing MW power. 
The studies in Refs. \cite{12,13} have revealed basic information about MW-induced Hall resistance, 
though the role of microscopic mechanisms still remains unclear.

Theoreticians, however, have started to work on MW-induced Hall resistance, suggesting several 
microscopic mechanisms that describe how $\Delta R_{xy}$ is affected by an ac field \cite{7,14}. 
Dissipative resistivity $\rho_{xx}(B)=\rho_{xx}(-B)$, whose change at low temperatures is 
governed mostly by the inelastic mechanism \cite{6,7}, remains an even function under magnetic 
field reversal. In contrast, the mechanisms responsible for Hall resistivity lead to both odd- 
and even-symmetry contributions in $\Delta R_{xy}$. The presence of even-symmetry terms was discussed 
\cite{7,15} in connection with the important question of violation of Onsager-Casimir relations 
\cite{16,17}. Indeed, since MW-excited electron systems are far from thermodynamic equilibrium 
conditions, it is quite possible that the symmetry of the resistivity tensor is essentially broken 
under MW irradiation. Further experimental investigations of MW-induced Hall resistance are 
desirable to gain more knowledge about mechanisms of MW photoresistance and related 
symmetry properties of resistivity.

In this work, we have carried out measurements of MW-induced Hall resistance in high-mobility 
bilayers formed in wide quantum wells (WQW's) with high electron density. Due to charge redistribution 
in WQW, there are two layers near the interfaces, separated by an electrostatic potential barrier, 
which creates a symmetric tunnel-coupled bilayer electron system with two populated 2D subbands 
closely spaced in energy \cite{10}. Despite a complex photoresponse in bilayer systems, the smaller 
period of MIS oscillations \cite{11} compared to the MIRO period permits us a direct visualization 
of the quantum component of magnetoresistance that is affected by microwaves. This fact might 
be considered as an experimental advantage compared to a 2DES with only one occupied subband 
\cite{18}. We show $\Delta R_{xx}$ and $\Delta R_{xy}$ for both directions of the perpendicular 
magnetic field $B$ and demonstrate that MW-induced Hall resistance exhibits an MIS/MIRO interference 
with a strong deviation from odd symmetry under field reversal. In addition, we find strong and 
non-trivial power and polarization dependences of $\Delta R_{xy}$. 

The paper is organized as follows. Section II presents experimental details on our samples and the 
experimental setup. Section III shows the results of photoresistance measurements including experiments 
where we have studied the dependence of photoresistance on the orientation of linear polarization. 
A discussion of the results in connection with theoretical models of MW-induced photoresistance 
in 2DES is presented in section IV. Concluding remarks are given in the final section.

\section{Experimental details}

Our samples are high-quality WQW's, see Ref. \cite{10}, with a well width of 45~nm, 
high electron density $n_{s}\simeq 9.1\times10^{11}$~cm$^{-2}$, and a mobility of 
$\mu~\simeq 1.9 \times 10^{6}$~cm$^{2}$/V s at $T=1.4$~K after a brief illumination 
with a red light-emitting diode. The samples have Hall-bar geometry (length $l~\times$ 
width $w$=500~$\mu$m$~\times~$200~$\mu$m) with six contacts. In our experiment, we have 
used both linear and indeterminate polarization (frequency range 35 to 170~GHz). 
Microwave irradiation is delivered in a circular waveguide down to the sample placed 
in a cryostat with a variable temperature insert. To control linear polarization 
of MW's, we employ special brass insets that reduce the transmission-line internal profile 
from the circular to a rectangular waveguide and vice versa. The insets are placed on both 
sides of the circular waveguide to control orientation of the MW field vector for linear 
polarization. In the case of indeterminate polarization, the inset close to the sample 
is replaced by a circular extension of the waveguide, which implies that we still have 
linear polarization (of the amount $\simeq$90~$\%$) but the orientation of the field vector is 
unknown. We measure MW-induced resistance in a single modulation (sm) technique and/or a
double-modulation (dm) technique for a direct measurement of photoresponse in order to 
improve the measurement resolution. The bias current is 1~$\mu$A. 
In the sm technique, the sample is exposed to a continuous MW irradiation and a voltage drop 
is measured between two voltage probes at a frequency of 13~Hz. In the dm technique, however, 
the MW's that are absorbed by the sample are amplitude modulated with an additional frequency of 333~Hz. This enables 
us to directly probe $\Delta R$. To probe symmetry of $\Delta R_{xx}$ and $\Delta R_{xy}$ 
under field reversal, we have used two samples in Hall bar geometry that demonstrate the 
best symmetry of MIS and Shubnikov-de Haas (SdH) oscillations for low- and high-field 
transport without MW excitation.

\section{Photoresistance measurements}

\begin{figure}[ht]
\includegraphics[width=9cm]{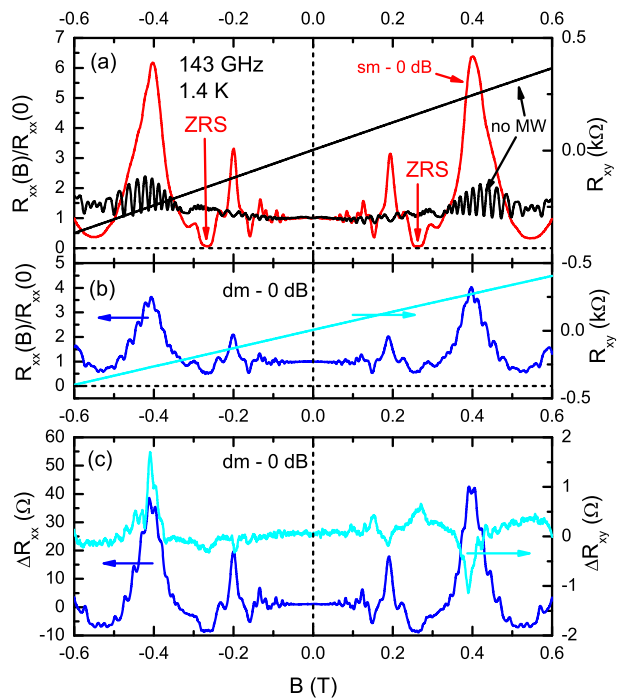}
\caption{\label{fig1} (Color online) (a) Normalized resistance $R_{xx}(B)/R_{xx}(0)$ under 
        microwave irradiation of 143~GHz at 1.4~K for 0~dB attenuation (sm technique)
        as well as dark magentoresistance (no MW) and Hall resistance $R_{xy}$. In $R_{xx}(B)/R_{xx}(0)$ we 
        observe a ZRS at $\pm$0.27~T. (b) $R_{xx}(B)/R_{xx}(0)$ and $R_{xy}$ for an 
        attenuation of 0~dB (dm technique). (c) Photoresistance $\Delta R_{xx}$ and 
        $\Delta R_{xy}$ measured in dm technique. We find an odd symmetry of $\Delta R_{xy}$ 
        under field reversal.}
\end{figure}

We start the presentation of our experimental results for a MW frequency of 143~GHz. 
In Fig. \ref{fig1}(a), we show first dark (no MW) normalized magnetoresistance 
$R_{xx}(B)/R_{xx}(0)$ and Hall resistance $R_{xy}$. The two-subband nature of our 
bilayer electron systems is confirmed by the presence of MIS oscillations, which occur 
for $|B|>0.1$~T and are superimposed on SdH oscillations at 1.4~K. If we apply a MW 
electric field with 0~dB attenuation at a frequency of 143~GHz in the sm technique, we 
observe ZRS for both negative and positive $B$ at $B$=$\pm$0.27~T \cite{10}.

The dissipative resistance in the dm technique does not show ZRS at $B$=$\pm$0.27~T 
owing to a loss in MW power due to the modulation, see Fig. \ref{fig1}(b). 
The difference is seen in a smaller amplitude of enhanced MIS oscillations, see the peaks at
$\pm 0.4$~T and $\pm 0.2$~T, and also confirmed by the appearance of SdH oscillations for $|B|> 0.4$~T. 
However, the electric field is strong enough to investigate the MW influence on our 2DES by a 
direct measurement of $\Delta R$ which will be used for further investigations. Photoresistance 
$\Delta R_{xx}$ and MW-induced Hall resistance $\Delta R_{xy}$ measured directly in the dm technique 
are shown in Fig. \ref{fig1}(c). The main features of the modified MIS oscillation pattern 
in $\Delta R_{xx}$, which is symmetric under field reversal, are seen also in 
$\Delta R_{xy}$ but whereas $\Delta R_{xx}$ obeys even symmetry under field reversal, we find 
odd symmetry in $\Delta R_{xy}$ (see the region of the enhanced MIS peak at $\pm$0.4~T and the ZRS 
region at $\pm$0.27~T). The result of odd symmetry in $\Delta R_{xy}$ is consistent with 
MW-induced Hall resistance in single-layer 2DES \cite{12,13}. However, having a closer 
look at $\Delta R_{xy}$ around $\pm$0.2~T, we find a weak feature which deviates from 
odd symmetry under field reversal. This warrants further investigation.

\begin{figure}[ht]
\includegraphics[width=9cm]{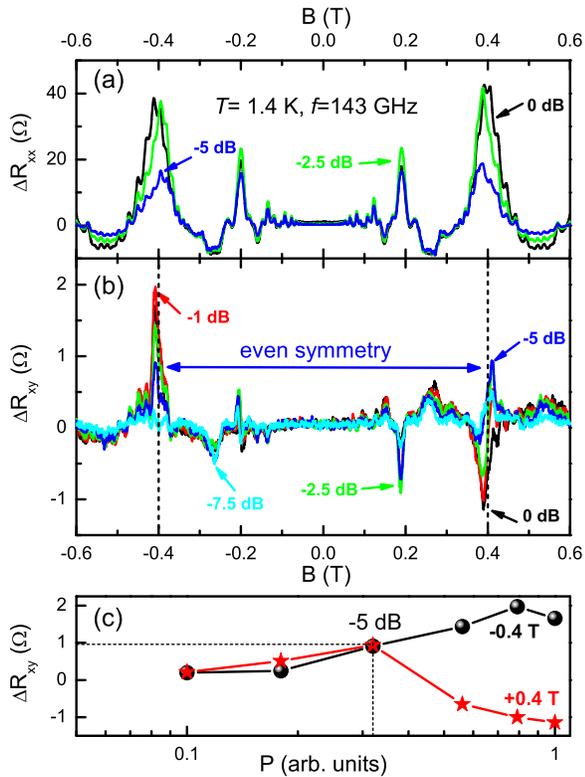}
\caption{\label{fig2} (Color online) (a) Power-dependent $\Delta R_{xx}$ and (b) MW-induced Hall resistance 
         at 143~GHz and 1.4~K. The enhanced resistance at +0.4~T changes sign leading to an even symmetry 
         for an attenuation of -5~dB. (c) $\Delta R_{xy}$ at $\pm$0.4~T as a function of MW power.}
\end{figure}

In Fig. \ref{fig2}, we illustrate the power dependence of photoresistance for $f$=143~GHz 
at 1.4~K. Starting again with $\Delta R_{xx}$ in Fig. \ref{fig2}(a), we find that the 
amplitude decreases with decreasing MW power (attenuation from 0 to -5~dB) while preserving
even symmetry under field reversal. However, the main peak of $\Delta R_{xy}$ at +0.4~T, see 
Fig. \ref{fig2}(b), changes its sign with decreasing MW power from 0 to $-7.5$~dB (in contrast 
to the peak at $-0.4$~T whose sign remains unchanged), so we find a transition from odd to even 
symmetry approximately at -5~dB. The Hall resistance (amplitude of the MIS peak) at $\pm 0.4$~T 
is plotted as a function of MW power in Fig. \ref{fig2}(c). In contrast to this change in symmetry 
at $\pm$0.4~T with decreasing MW power, we notice that all other features in $\Delta R_{xy}$, 
e.g. at $\pm$0.2~T and $\pm$0.27~T, do not exhibit an apparent change in symmetry. In addition, 
we find that with increasing temperature from 1.4 to 4~K (not shown here), only the amplitude of MIS 
oscillations in $\Delta R_{xy}$ decreases whereas the symmetry of $\Delta R_{xy}$ is preserved. 
This observation can be considered as an indication of the temperature independence of microscopic 
mechanisms contributing to MW-induced Hall resistance.

\begin{figure}[ht]
\includegraphics[width=9cm]{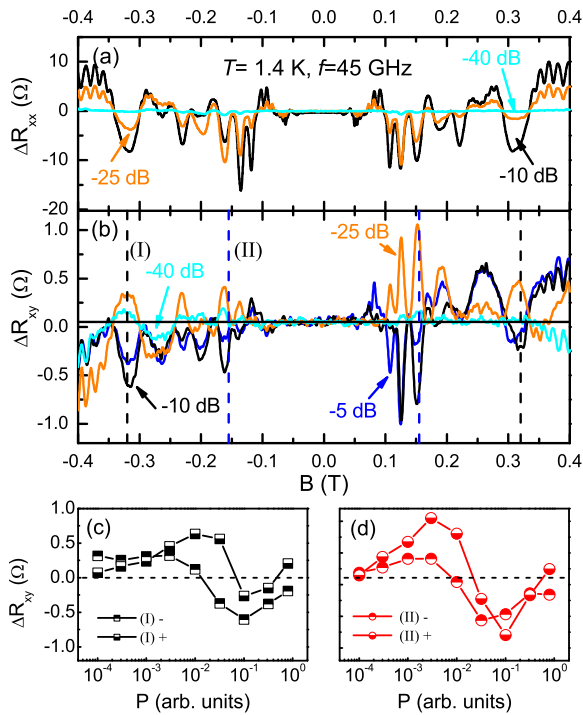}
\caption{\label{fig3} (Color online) (a) Power dependence of $\Delta R_{xx}$ and (b) $\Delta R_{xy}$ 
         for different chosen attenuations at 45~GHz and 1.4~K. For that frequency, $\Delta R_{xy}$ 
         exhibits power-dependent MW-induced Hall resistance, denoted with peaks (I) and (II), which change 
         their sign giving rise to odd or even symmetry. $\Delta R_{xy}$ in peak (I) (c) and in peak (II) 
         (d) for $\pm B$ as a function of MW power.}
\end{figure}

Whereas the inversion of \textit{one} particular feature in Hall resistance leading to breaking of 
odd symmetry under field reversal has been observed at high frequencies, numerous power-dependent 
measurements of $\Delta R_{xy}$ for $f<75$~GHz demonstrated that several features in $\Delta R_{xy}$ 
for both negative and positive magnetic field change their signs in a non-trivial way as power varies. 
As an example of this behavior, we present the power dependence of both $\Delta R_{xx}$ and $\Delta R_{xy}$ 
for 45~GHz at 1.4~K in Fig. \ref{fig3}. Due to the possibility of our microwave setup, we investigate 
photoresistance at elevated MW power (see the estimates of the MW electric field in Sec. IV).

As a first impression, in Fig. \ref{fig3}(b) we see several MIS oscillations 
changing sign with decreasing MW power. These main features correlate with the
MW response in $\Delta R_{xx}$, see Fig. \ref{fig3}(a). It should be noticed that 
for lower MW frequencies, see also Ref. \cite{8}, MIS oscillations show a more
complicated behavior compared to high frequencies where several MIS peaks are strongly 
enhanced but they can be successfully described by the model used in Ref. \cite{8}.
Nevertheless, we get a reasonable even symmetry in $\Delta R_{xx}$ under field reversal 
although MIS oscillations around $\pm 0.16$~T differ in amplitude. Let us now focus on 
MW response in $\Delta R_{xy}$ in Fig. \ref{fig3}(b). For a better analysis, we mark 
two MIS oscillation peaks in $\Delta R_{xy}$ for $\pm B$ in Fig. \ref{fig3}(b) and denote 
them as (I) and (II). A change in sign of MW-induced Hall resistance occurs with 
decreasing MW power (from -5 to -40~dB) for all peaks except the one at $\pm 0.26$~T, 
where odd symmetry persists with changing attenuation. In particular, for $0<B< 0.15$~T 
we observe several MIS oscillations whose amplitude is strongly enhanced by microwaves 
and which are inverted with decreasing MW power (compare data for -5 and -25~dB attenuation). 
However, such a behavior with a comparable amplitude of MIS peaks is not observed for 
$-0.15 <B<0 $~T. In summary, Hall resistance at 45~GHz excitation obeys neither odd nor 
even symmetry. We now look more closely at the peaks marked with (I) and (II), for which we 
plot $\Delta R_{xy}$ as a function of MW power in Fig. \ref{fig3}(c) and (d). The equal sign 
of $\Delta R_{xy}$ for positive and negative $B$ can be considered as an indication of 
even symmetry with respect to field reversal. Starting with peak (I) at $\pm 0.32$~T, 
see Fig. \ref{fig3}(c), we always find even symmetry except for an attenuation of -1~dB 
and around -15~dB. For peak (II) at $\pm 0.155$~T, we see a similar behavior, i.e. even 
symmetry is preserved with changing MW attenuation, except for a narrow region where 
peak flips occur. It is worth noting that all peak flips in MW-induced Hall resistance 
appear in the regions of MW power and magnetic field where MW's strongly affect the 
photoresistance $\Delta R_{xx}$.

\begin{figure}[ht]
\includegraphics[width=9cm]{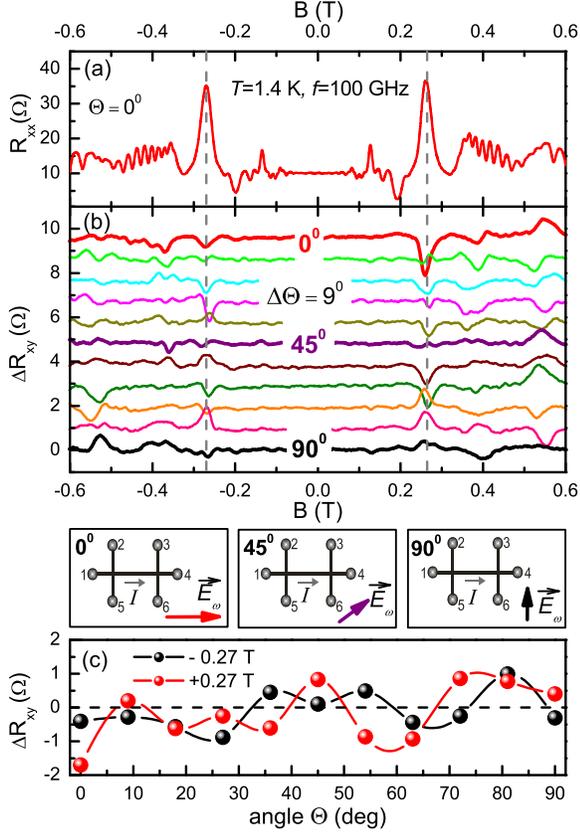}
\caption{\label{fig4} (Color online) (a) $R_{xx}$ for $\Theta=0^{\circ}$ and (b) MW-induced Hall 
         resistance $\Delta R_{xy}$ for 100~GHz at 1.4~K dependent
         on the orientation of linear polarization from $\Theta=0^{\circ}$ to $\Theta=90^{\circ}$, in steps of
         $\Delta\Theta=9^{\circ}$. (c) $\Delta R_{xy}$ at $\pm 0.27$~T as a function
         of orientation of the electric field (tilt angle with respect to current direction - see sketches) 
         exhibits alternating sign leading to odd (e.g. $\Theta=54^{\circ}$,$\Theta=90^{\circ}$) and even 
         (e.g. $\Theta=18^{\circ}$, $\Theta=81^{\circ}$) symmetry. Orientation of the vector ${\bf E}_{\omega}$ 
         is depicted in sketches for $\Theta=0^{\circ}$, $\Theta=45^{\circ}$ and $\Theta=90^{\circ}$.}
\end{figure}

We have also performed measurements where we control the linear polarization, i.e.' the 
orientation of the MW electric field vector ${\bf E}_{\omega}$. Due to a loss in MW power 
with brass insets, we carry out measurements in the sm technique and extract $\Delta R_{xy}$. 
We focus here on an intermediate frequency of $f$=100~GHz. This frequency has been chosen 
because (i) the rectangular output of the brass inset is large enough to ensure a high 
enough MW electric field estimated to $E_{\omega}\simeq 1.5$~V/cm for 0~dB attenuation 
(this is not the case for $f>$100~GHz due to other brass insets), and (ii) a complicated 
power dependence for a fixed polarization, as for the case of 45~GHz, is avoided. Another argument 
in favour of $f$=100~GHz is the presence of one strongly enhanced MIS peak in $R_{xx}$ at 
$\pm$0.27~T and the corresponding prominent feature in $\Delta R_{xy}$, see Fig. \ref{fig4}(a,b), 
whose behavior is convenient to follow. The orientation of the electric field, i.e. tilt angle 
with respect to current direction is also sketched for the angles $\Theta=0^{\circ}$, 
$\Theta=45^{\circ}$ and $\Theta=90^{\circ}$ in Fig. \ref{fig4}. MW-induced Hall resistance 
$\Delta R_{xy}$ is measured at the highest possible MW power (close to 0~dB attenuation) 
at 1.4~K, shown in steps of $\Delta\Theta=9^{\circ}$ in Fig. \ref{fig4}(b).  
To ensure the same MW electric field for all the tilt angles, we have compared the 
amplitude of SdH oscillations in $R_{xx}$ under the same conditions and found that 
it remains constant \cite{8}. Whereas $R_{xx}$ (and thus $\Delta R_{xx}$) does not 
depend on linear polarization \cite{8,9,10,18}, MW-induced Hall resistance exhibits 
essential angular dependence. If we focus on the enhanced peak in $R_{xx}$ at $\pm 0.27$~T, 
$\Delta R_{xy}$ shows even symmetry for e.g. $\Theta=18^{\circ}$ and odd symmetry is 
observed for e.g. $\Theta=54^{\circ}$. This is also illustrated in Fig. \ref{fig4}(c), 
where we plot the amplitude of $\Delta R_{xy}$ as a function of the angle between the current 
${\bf I}$ and electric field ${\bf E}_{\omega}$ indicating a somehow oscillating behavior 
with increasing tilt angle. This result strongly indicates that in contrast to $\Delta R_{xx}$, 
$\Delta R_{xy}$ is sensitive to linear polarization of incident MW radiation and, consequently, 
microscopic mechanisms that account for $\Delta R_{xy}$ depend on the orientation of linear 
polarization.

\section{Discussion}

Our experiments have undoubtedly shown that MW irradiation affects Hall resistance depending 
on MW power and orientation of linear polarization. Compared to power-dependent $\Delta R_{xy}$ 
in single-layer systems with a mobility of $1.5 \times 10^7$ cm$^2$/V s, see Ref. \cite{13}, where a 
progressively stronger modulation of MIRO's is observed with increasing MW intensity, we 
demonstrate that our systems behave differently. The Hall resistance oscillations neither 
show that progressive increase with power, nor obey the odd symmetry observed in Ref. \cite{13}.

In that context, we first discuss existing theoretical models based on bulk mechanisms of 
photoconductivity \cite{7,14,15}. Considering a 2DES in the $(xy)$ plane, and assuming a linear 
regime of dc response to the applied field ${\bf E}$, one can write a conventional expression 
for the current density, ${\bf j}= \hat{\sigma} {\bf E}$, where the conductivity tensor in 
the $(xy)$ space is written as 
\begin{equation}
\hat{\sigma}= \left( \begin{array}{cc} \sigma_D + \delta \sigma_{Ds}+\delta \sigma_{Da} & -\sigma_H + 
\delta \sigma_{Hs}+\delta \sigma_{Ha} \\ \sigma_H + \delta \sigma_{Hs}-\delta \sigma_{Ha} & 
\sigma_D + \delta \sigma_{Ds}-\delta \sigma_{Da} \end{array} \right),
\label{eq1}
\end{equation}
$\sigma_D$ and $\sigma_H$ are the dissipative and Hall conductivities in the absence of MW excitation, 
while $\delta \sigma_{Di}$ and $\delta \sigma_{Hi}$ are the symmetric ($i=s$) and antisymmetric 
($i=a$) MW-induced contributions to these conductivities. The resistivity tensor $\hat{\rho}$, 
defined according to ${\bf E}= \hat{\rho} {\bf j}$, is obtained directly from Eq. (1), and 
its non-diagonal component $\rho_{xy}$ is equal to the Hall resistance $R_{xy}$:
\begin{equation}
R_{xy}= \frac{\sigma_H - \delta \sigma_{Hs}-\delta \sigma_{Ha}}{(\sigma_H - \delta \sigma_{Ha})^2+
(\sigma_D + \delta \sigma_{Ds})^2 - \delta \sigma_{Hs}^2-\delta \sigma_{Da}^2 }. 
\label{eq2}
\end{equation}
Below the onset of the quantum Hall effect, the dark Hall resistance $R_{xy}^{(0)}=
\sigma_H/(\sigma_H^2+\sigma_D^2)$ is very close to the classical Hall resistance $R_H=B/en_s$.
  
According to Refs. \cite{6,7}, the main photoinduced contribution to the dissipative conductivity 
is $\delta \sigma_{Hs}$, caused (at low temperatures) mostly by the inelastic mechanism. The 
contributions $\delta \sigma_{Da}$, $\delta \sigma_{Hs}$, and $\delta \sigma_{Ha}$ are determined 
by the other three mechanisms called the displacement, the photovoltaic, and the quadrupole 
ones \cite{7,15}. Since these contributions are much smaller than $\sigma_D+\delta \sigma_{Ds}$, 
one can rewrite Eq. (2) as 
\begin{eqnarray}
R_{xy}=R_H+\rho^{(1)}_{xy}+\rho^{(2)}_{xy},~~\rho^{(1)}_{xy}=
\frac{\delta \sigma_{Ha}-\delta \sigma_{Hs}}{\sigma_H^2}, \nonumber \\ 
\rho^{(2)}_{xy}= -\frac{(\sigma_{D}+\delta \sigma_{Ds})^2}{\sigma_H^3},
\label{eq3}
\end{eqnarray}
where it is also taken into account that $\sigma_D+\delta \sigma_{Ds} \ll \sigma_H$ and 
$\sigma_H \simeq 1/R_H$, which assumes a finite (not very small) magnetic field. 

Thus, the MW-induced modification of Hall resistance is given by two terms. The first 
one is determined directly by the Hall photoconductivity contributions $\delta \sigma_{Ha}$ 
and $\delta \sigma_{Hs}$, while the second one is related to dissipative 
resistivity. The theory \cite{7,15} attributes $\delta \sigma_{Ha}$ and $\delta \sigma_{Hs}$ 
to the contributions of the photovoltaic and quadrupole mechanisms, respectively. 
Both of these contributions retain odd symmetry under magnetic field reversal, though 
the presence of $\delta \sigma_{Hs}$ leads to a violation of the relation $\sigma_{xy}(B)=
\sigma_{yx}(-B)$. The nature of the quadrupole mechanism suggests that $\delta \sigma_{Hs}$ 
is polarization-dependent. However, the consideration in Refs. \cite{7,15} 
is valid for elliptical polarization of a MW field in the main axes ($xy$) and can be applied 
for linear polarization along either the $x$ or $y$ axis ($\Theta=0^{\circ}$ or $\Theta=90^{\circ}$). 
The case of arbitrary linear polarization is studied in Ref. \cite{14} by considering the 
displacement mechanism of photoconductivity. It was shown that the corresponding 
$\delta \sigma_{Hs}$ contains a contribution that is an even function of the magnetic 
field. This contribution depends on the tilt angle as $\sin(2 \Theta)$. In summary, the 
term $\rho^{(1)}_{xy}$ can lead to an even-symmetry part of the Hall resistivity tensor, 
and this part is polarization-dependent.

In contrast, the term $\rho^{(2)}_{xy}$ possesses an odd symmetry under field reversal and 
is polarization-independent. As a first impression, this term should be less significant, 
because of strong inequality $\sigma_D+\delta \sigma_{Ds} \ll \sigma_H$. However, another 
strong inequality, $|\delta \sigma_{Ds}| \gg |\delta \sigma_{Ha} \pm \delta \sigma_{Hs}|$, 
appears to be more important, and our estimates prove that $\rho^{(2)}_{xy}$ is the main 
part of the Hall resistance under our experimental conditions. Specifically, we 
have applied theoretical expressions \cite{7} for $\delta \sigma_{Ha}$ and $\delta \sigma_{Hs}$ 
with known MW field $E_{\omega} \simeq 2$~V/cm (0~dB attenuation) for $f=143$~GHz and 
$E_{\omega} \simeq 3$~V/cm (-10~dB attenuation) for $f=45$~GHz, and we found that  
$\rho^{(1)}_{xy}$ is approximately one order of magnitude smaller than 
$\rho^{(2)}_{xy}$ (and even smaller for $f=143$~GHz).
A similar conclusion is made in Ref. \cite{14} by proving that at $\Theta=45^{\circ}$, 
when the even-symmetry part of the Hall resistivity tensor is maximal, this part still 
does not produce an appreciable deviation from the odd symmetry. Therefore, the main 
contribution to the Hall resistivity comes from $\rho^{(2)}_{xy}$ and the odd symmetry
should be preserved. The importance of the term $\rho^{(2)}_{xy}$ is also emphasized in 
Ref. \cite{19}. 

\begin{figure}[ht]
\includegraphics[width=9cm]{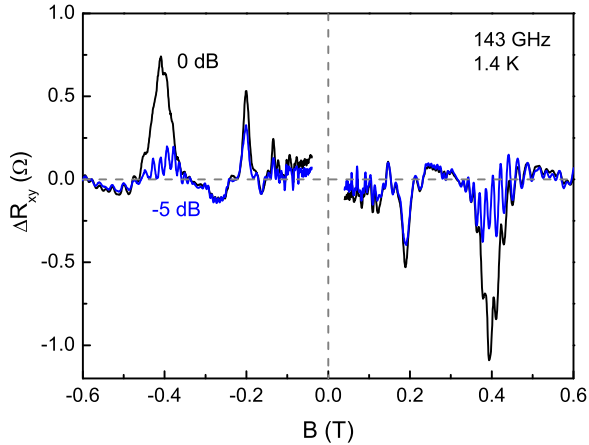}
\caption{\label{fig5} (Color online) MW-induced Hall resistance $\Delta R_{xy}$ for 143~GHz 
(0 dB and -5 dB attenuation) and 1.4~K calculated from Eq. (4).}
\end{figure}

By retaining only $\rho^{(2)}_{xy}$ in Eq. (3), one can rewrite the MW-induced  
Hall resistivity in the form 
\begin{equation}
\Delta R_{xy} \simeq \frac{\rho_{xx}^{(0)2}-\rho_{xx}^2}{R_H}, 
\label{eq4}
\end{equation}
where $\rho_{xx}^{(0)}$ is the dark dissipative resistivity. This relation 
does not contain the details of photoconductivity mechanisms and can be checked 
directly by using the dissipative resistivity $\rho_{xx}$ measured in the experiment. 
We have carried out such a procedure for both 143 and 45 GHz and different attenuations. 
The results for 143 GHz are presented in Fig. \ref{fig5}. One can see that the amplitudes of 
the main peaks at $\pm 0.2$~T and $\pm 0.4$~T are close to those obtained experimentally 
and behave in a similar way with decreasing power. The sign inversion of $\Delta R_{xy}$ 
just above +0.4~T, leading to apparent reversal of symmetry from odd to even one for -5 dB 
attenuation, can be explained in terms of a slight asymmetry of measured dissipative 
resistivity [see Fig. \ref{fig2}(a)] and alteration of the sign of $\rho_{xx}^{(0)2}-\rho_{xx}^2$ 
in this region of magnetic field.  

The amplitudes of the peaks in $\Delta R_{xy}$ for other MW frequencies used in the 
experiment are also in reasonable agreement with Eq. (\ref{eq4}). However, neither the 
strong and complicated modifications of the Hall resistance with MW power at 45 GHz (Fig. \ref{fig3}) 
nor the polarization dependence (Fig. \ref{fig4}) can be reproduced using this simple expression.
In this connection, we again discuss the possible influence of the term $\rho^{(1)}_{xy}$, 
which is essentially determined by the Hall photoconductivity mechanisms. In principle, 
this term can become comparable with $\rho^{(2)}_{xy}$ if the field $E_{\omega}$ is 
considerably larger than that used in our estimates. A theoretical study in Ref. \cite{20} 
suggests that the MW field in the near-contact regions of the Hall bar is strongly enhanced 
compared to the MW field in the bulk of the sample. Therefore, the possibility that 
electrons feel a stronger field should not be disregarded. The Hall photoconductivity 
mechanisms can lead to inversion of oscillation peaks in $\rho^{(1)}_{xy}$ under a transition
to the regime when the oscillating non-equilibrium part of electron distribution saturates 
with increasing $E_{\omega}$. This saturation effect is discussed in detail in 
Refs. \cite{6,7}. However, our estimates show that at 45~GHz and -10~dB attenuation 
($E_{\omega} \simeq 3$~V/cm), the 2DEG at 1.4~K is already in the saturation regime. The assumed 
increase of $E_{\omega}$ in the near-contact regions cannot, therefore, cause inversion 
of oscillations in $\Delta R_{xy}$ due to the saturation effect. Furthermore, if we look 
at the polarization dependence, the even contribution to the Hall resistivity in 
$\rho^{(1)}_{xy}$ should follow, according to the theory, a simple $\sin(2 \Theta)$ 
law. Instead, in Fig. \ref{fig4} we see multiple oscillations of both even and odd contributions 
as functions of $\Theta$, and the even contribution does not disappear at $\Theta=0^{\circ}$ 
and $\Theta=90^{\circ}$, contrary to the theoretical prediction. Therefore, even if we assume 
that the field $E_{\omega}$ is effectively enhanced, the complicated behavior of the observed 
MW-induced Hall resistance [Figs. \ref{fig3} and \ref{fig4}] cannot be explained by employing 
the bulk mechanisms of MW photoconductivity.

An alternative approach to the MW-induced effects in dissipative resistance such as MIRO 
and ZRS was recently proposed in Ref. \cite{21}. It is suggested that these effects have 
a purely classical origin. They are induced by ponderomotive forces that arise in the 
near-contact regions because of a strong inhomogeneity of the MW field and possess an oscillatory 
dependence on MW frequency and magnetic field. It is not clear, however, whether the 
presence of such ponderomotive forces can contribute to the Hall resistance $\Delta R_{xy}$. 
In any case, it cannot lead to a polarization dependence of $\Delta R_{xy}$, because the MW 
polarization in the near-contact regions is fixed (the MW field is perpendicular to the boundary 
between 2DEG and contact) regardless of polarization of the incident wave. The non-trivial 
power dependence of $\Delta R_{xy}$ observed in our experiment cannot be explained within 
this approach as well. 

Another approach to the MW-induced magnetotransport is developed in Ref. \cite{22}, where
the influence of MW's on the edge trajectories has been studied and the appearance of ZRS 
is explained in terms of stabilization of the edge-state transport by MW's. A deviation of $R_{xy}$ 
from the classical Hall resistance, which correlates with the corresponding changes in the 
dissipative resistance $R_{xx}$, is also mentioned. Both $R_{xx}$ and $R_{xy}$ are found to 
be sensitive to the direction of linear polarization of MW's. The implication of these results 
to symmetry properties of $R_{xy}$ with respect to magnetic field reversal has not been 
discussed. The theory of Ref. \cite{22} might be relevant to the samples with very high 
mobilities, such as those studied in Ref. \cite{13}, where edge trajectories are still important 
for transport in the region of magnetic fields below 0.5~T. Presumably, the edge-state 
transport in these samples is responsible for the fact that the oscillations of MW-induced 
Hall resistance $\Delta R_{xy}$ and the oscillations of $\rho_{xx}$ have comparable amplitudes, 
which also means that $\Delta R_{xy}$ cannot be described by Eq. (\ref{eq4}). In our samples, however, 
the transport is expected to be bulk-like (diffusive) in the mentioned region of magnetic fields, 
since Eq. (\ref{eq4}) proves to be applicable for estimating the magnitude of MW-induced $\Delta R_{xy}$.  

Finally, we cannot completely rule out the possibility that the complicated behavior of 
$\Delta R_{xy}$ is related to specific features of bilayer (two-subband) systems as compared 
to single-layer systems. Above, we have stated that the only essential difference between 
magnetoresistances of single-subband and two-subband systems is the modulation of the 
quantum contribution to resistivity by the MIS oscillations in two-subband 2DES. This 
statement is well justified from the point of view of bulk transport theory and is confirmed 
in numerous experiments \cite{8,9,10,11,18,23,24,25}. 
In addition, the theoretical model of dissipative MW photoresistance based on a consideration of 
the inelastic mechanism \cite{6} explains satisfactorily all features of MIRO's (including frequency, 
power, and temperature dependence) for different two-subband systems studied in our experiments 
(see, e.g. Refs. \cite{8} and \cite{10}). The MW-induced Hall resistance, however, is a subtle 
effect: in our sample, $\Delta R_{xy} \ll \Delta R_{xx}$. If the transport is influenced 
by the presence of sample edges or contact regions, the bilayer nature of our system may 
essentially manifest itself in $R_{xy}$. To check out this assumption, it is desirable to 
measure MW-induced $R_{xy}$ in single-layer 2DES whose density and mobility are close to 
those of our system.   

\section{Conclusions}

We have studied the photoresponse of $\Delta R_{xx}$ and $\Delta R_{xy}$ in 
high-quality bilayer electron systems. Whereas even symmetry is preserved in $\Delta R_{xx}$ 
with a reasonable accuracy, we found a violation of odd symmetry in MW-induced Hall resistance, 
in contrast to previous experiments \cite{12,13} on single-layer 2DES with higher mobilities. 
A non-trivial power dependence is observed for several MIS oscillation peaks in $\Delta R_{xy}$. 
Symmetry of $\Delta R_{xy}$ is also essentially modified by changing MW power. Varying 
$\Delta R_{xy}$ for different orientations of linear polarization strongly confirms the feasibilitly 
of polarization-dependent microscopic mechanisms of MW-induced Hall resistance, in contrast to 
polarization immunity in dissipative resistance \cite{8,10,26}. The photoresponse in 
$\Delta R_{xy}$ might be accounted for by the presence of two components in $\Delta R_{xy}$, of 
which one is odd and another is even with respect to magnetic field reversal. A reasonably 
good estimate for the magnitude of $\Delta R_{xy}$ is obtained within the bulk transport approach. 
However, bulk transport models, as well as currently existing alternative approaches to the 
problem of MW-induced resistance, fail to explain non-trivial power and polarization dependence 
and the strong violation of odd symmetry observed in our experiments. Due to the essential 
deviation of our data from the theoretical models discussed above, we cannot preclude the influence 
of possible previously unconsidered microscopic mechanisms that might exist in a broad interval 
of MW power or turn on at elevated MW power. 

In general, our data and its analysis suggest that the problem of MW-induced Hall resistance 
(and, hence, the related problem of MW-induced dissipative resistance) still remains a puzzle 
that awaits a future solution. We suppose that a theory that could describe the behavior of 
both components of the MW-induced resistivity on an equal footing and explain the variety of 
experimental facts has to be based on a consideration of quantum transport in the presence of 
a strongly inhomogeneous MW field. We assume that our systematic study will stimulate further 
experimental and theoretical investigations, which are crucial to clarify the origin of 
MW-induced phenomena in 2DES. 

We acknowledge support from COFECUB-USP (project number U$_{c}$ Ph 109/08), FAPESP and CNPq
(Brazilian agencies). For measurements we have used microwave facilities from ANR MICONANO.

\end{document}